\documentclass[11pt]{article}
\usepackage{amssymb,cite,graphics}
\newcommand\toline[1]{--#1}
\newcommand{\fract}[2]{{\textstyle\frac{#1}{#2}}}
\newcommand{\fr}[2]{{\textstyle\frac{#1}{#2}}}
\newcommand{\ri}{\right}

\newcommand{\lf}{\left}
\newcommand{\te}{\theta}
\newcommand{\CS}{{\cal S}}

\newcommand{\CA}{{\cal A}}
\newcommand{\CN}{{\cal N}}
\newcommand\eq{\begin{equation}}
\newcommand\en{\end{equation}}
\newcommand\bea{\begin{eqnarray}}
\newcommand\eea{\end{eqnarray}}
\newcommand\nn{\nonumber}

\newcommand{\resection}[1]{\setcounter{equation}{0}\section{#1}}

\begin{document}
\begin{titlepage}
\vskip 0.5cm
\begin{flushright}
DTP-99/69 \\
ITFA 99-27 \\
{\tt hep-th/9910102}
\end{flushright}
\vskip 1.5cm
\begin{center}
{\Large{\bf
Differential equations and integrable models:\\[5pt]
the $SU(3)$ case
}}
\end{center}
\vskip 0.8cm
\centerline{Patrick Dorey%
\footnote{e-mail: {\tt p.e.dorey@durham.ac.uk}}
and Roberto Tateo%
\footnote{e-mail: {\tt tateo@wins.uva.nl}}}
\vskip 0.9cm
\centerline{${}^1$\sl\small Dept.~of Mathematical Sciences,
University of Durham, Durham DH1 3LE, UK\,}
\vskip 0.2cm
\centerline{${}^2$\sl\small Universiteit 
van Amsterdam, Inst.~voor Theoretische
Fysica, 1018 XE Amsterdam, NL\,}
\vskip 1.25cm
\begin{abstract}
\noindent
We exhibit a relationship between
the massless $a_2^{(2)}$ integrable quantum field theory
and a certain third-order
ordinary differential equation, thereby extending 
a recent  result
connecting the massless sine-Gordon
model to the Schr\"odinger equation.
This forms  part of a more general
correspondence involving 
$A_2$-related Bethe ansatz systems
and  third-order differential equations.
A non-linear integral equation for the generalised spectral problem
is derived, and some numerical checks are performed.
Duality properties are discussed, and
a simple  variant of the nonlinear  equation is suggested as a
candidate to describe the finite volume ground state energies 
of minimal conformal field theories perturbed by the operators $\phi_{12}$,
$\phi_{21}$ and $\phi_{15}$. This is
checked against previous results obtained using
the thermodynamic Bethe ansatz.\\


\end{abstract}
\end{titlepage}
\setcounter{footnote}{0}
%%%%%%%%%%%%%%%%%%%%%%%%%%%%%%%%%%%%%
\def\thefootnote{\fnsymbol{footnote}}
%%%%%%%%%%%%%%%%%%%%%%%%%%%%%%%%%%%%%%%%%%%%%%%%%%%%%%%%%%%%%%%%%%%%%
%%%  start of the letter %%%%%%%%%%%%%%%%%%%%%%%%%%%%%%%%%%%%%%%%%%%%
%%%%%%%%%%%%%%%%%%%%%%%%%%%%%%%%%%%%%%%%%%%%%%%%%%%%%%%%%%%%%%%%%%%%%
%
\resection{Introduction}
\label{intr}
A curious
connection between certain integrable quantum field
theories and the theory of the Schr\"odinger equation has been the 
subject of some recent work~\cite{DT3,BLZschr,Suz1,DT4}.
In this paper we 
extend these results
by establishing a link between
functional relations for
$A_2$-related Bethe ansatz systems (see for example~\cite{BR,PS})
and third-order differential equations. 
Most of our analysis concerns a certain 
specialisation of the model, a particularly symmetric case
that can  also be related to the dilute A-model of~\cite{WNS}.

In the cases studied in \cite{DT3,BLZschr,Suz1,DT4},
the most general differential equation was a
radial Schr\"odinger problem with `angular momentum' $l$ and 
homogeneous potential $x^{2M}$, initially defined
on the positive real axis $x\in(0,\infty)\,$:
\eq
\lf ( - {d^2 \over dx^{2}} + x^{2M} + {l(l+1) \over  x^{2}}  \ri )
\psi(x,E) = E \psi(x,E)~.
\label{Sch}
\en
The relevant integrable quantum field theories were 
the massless twisted sine-Gordon models or, equivalently,  
the twisted XXZ/6-vertex  models in their thermodynamic limits, and
their reductions. It is worth noting that these models are
all related to the Lie algebra $A_1$.
Spectral functions 
associated with~(\ref{Sch}) satisfy  functional 
relations~\cite{Sib,V0},
and these were mapped into functional equations appearing
in the context of integrable quantum field 
theory in~\cite{DT3,BLZschr,Suz1,DT4}. We will follow a similar
strategy here, taking a simple third-order ordinary differential
equation as our starting-point and showing that the Stokes
multipliers and certain spectral functions
for this equation together satisfy relations which are essentially
the analogues, for the 
Bethe ansatz systems treated in
\cite{PS,WNS},
of the T-Q systems which arise in the context of
the integrable quantum field theories related to 
$A_1$~\cite{Bax1,BLZ2}.
This is the subject of \S2, while in \S3 we borrow some other ideas
from integrable quantum field theory in order to derive a nonlinear
integral equation for the spectral functions, an equation which is put
to the test in a simple example in \S4.
Duality properties are discussed in \S5, allowing us to find the 
equivalent of the angular-momentum term in (\ref{Sch}) for the 
third-order equation. Connections with various perturbed conformal
field theories are discussed and tested in \S6. Finally \S7 discusses
the
most general $A_2$-related BA equations that arise in this context,
and \S8 contains our conclusions.
\resection{The differential equation}
\label{sdifeq}
We begin with the following
third-order ordinary differential equation:
\eq
y'''(x,E) + P(x,E) y(x,E)=0~,
\label{thirdo}
\en
and initially restrict ourselves to purely homogeneous `potentials'
$x^{3M}$, giving  $P(x,E)$
the form 
\eq
P(x,E)= x^{3M}  - E~.
\label{pot}
\en
These are the simplest higher-order
generalisations of
the $l=0$ cases of (\ref{Sch}), and so
we expect that some of the 
properties of that equation,
used in the 
analysis of~\cite{DT4},
will be preserved. In particular, motivated by
the results of \cite{Sib} for second-order equations,
we suppose that
(\ref{thirdo}) has a 
solution $y=y(x,E)$ such that: \\
(i) $y$ is an entire function of $(x,E)$ though, due to the branch
point in
the potential at $x=0$, $x$ must in 
general be considered to live on a suitable cover of 
the punctured complex plane; \\
(ii) $y$ ,  $y'= dy/dx$ and $y''= d^2y/dx^2$ admit, for 
$M>1/2$,  the asymptotic representations
\eq
y    \sim    x^{-M} e^{ -\fract{1}{M+1} x^{M+1}}~~,~~
y' \sim   - e^{ -\fract{1}{M+1} x^{M+1}}~~,~~        
y''  \sim   x^{M}  e^{ -\fract{1}{M+1} x^{M+1}}~~, 
\label{largex}
\en
as $x$ tends to infinity in the sector
\eq
|\arg x\,|<\frac{4 \pi}{3M+3}~.
\label{sector}
\en
\noindent
Furthermore, 
these asymptotics, or even just the asymptotic of
$y(x,E)$ with $x$ remaining on the
positive real axis, characterise 
$y$ uniquely.

For $M \le 1/2$, the story is complicated by the appearance of extra
terms in the asymptotic (\ref{largex}). 
The behaviour of the solution which decays as $x\rightarrow+\infty$
can be more generally found from the formula
\eq
y(x,E)\sim
P(x,E)^{-1/3}\exp(-\int^x_{x_0} P(t,E)^{1/3}dt)\,,
\label{secla}
\en
with the constant $x_0$ being related to the normalisation of the
solution.  (This is the analogue of an
approximate WKB solution of a  Schr\"odinger equation.)
Since to take $M\le 1/2$ would bring other technical problems into the
treatment to be given below,
from now on, unless otherwise stated, we shall restrict ourselves to
 $M>1/2$.  This range is 
the analogue of the `semiclassical domain'
of \cite{BLZ2} (see \cite{DT3,BLZschr,DT4} for a discussion in the
context of differential equations).

Given $y(x,E)$, bases of solutions to
the third-order equation can be constructed just as in the
second-order case. For general values of $k$, define 
\eq
y_k(x,E)= 
\omega^k y(\omega^{-k} x, \omega^{-3 M k }E)~,
\label{ykdef}
\en
with 
\eq
\omega= \exp \lf({2\pi i \over 3 M+3} \ri)~.
\en
Then $y_k$ solves
\eq
y'''_k(x,E) +  e^{-2k\pi i} P(x,E)
 y_k(x,E)=0~,
\label{thirdotr}
\en
and so when $k$ is an integer it provides a
(possibly new) solution
to the original problem~(\ref{thirdo}). However,
since we will shortly need to
consider fractional values, we will leave $k$ arbitrary for
now.
It is convenient to
define sectors $\CS_k$ as
\eq
\CS_k~:~~~ 
\lf| \arg x -\frac{2k\pi}{3M{+}3}\ri| < \frac{\pi}{3M{+}3}~.
\en
On the cover of the punctured complex plane on which $x$ is defined,
the sector $\CS_k$ abuts the sectors $\CS_{k-1}$ and $\CS_{k+1}$, and
the sector (\ref{sector}) is 
$\CS_{-3/2}\cup\CS_{-1/2}\cup\CS_{1/2}\cup\CS_{3/2}$. 
The pattern of dominance and subdominance of solutions is 
more involved than in the second-order case, since there are now 
three different behaviours for solutions at large $|x|$.
In addition to a solution
with leading behaviour
$x^{-M}\exp(-x^{M+1}/(M{+}1))$ as $|x|\rightarrow +\infty$,
there are also solutions which behave as $x^{-M}\exp(e^{\pm\pi
i/3}x^{M+1}/(M{+}1))$. (This is simply a consequence of the fact that
the three third roots of $-1$ are $-1$, $e^{\pi i/3}$ and
$e^{-\pi i/3}$.) Depending on the sector, either one or two of these
solutions tend to zero at large $|x|$. We call `subdominant' the
solution which tends to zero {\em fastest} in a given sector; then, up
to a scalar multiple,
$y_k$ is characterised as
the unique solution to (\ref{thirdotr}) subdominant inside
$\CS_k$.

The asymptotic (\ref{largex}) and the definition
(\ref{ykdef}) together imply
\bea
y_k    \sim   \omega^{(M{+}1)k} x^{-M}
 e^{ -\fract{1}{M+1}\omega^{-(M{+}1)k} x^{M+1}}~,&&~
y_k' \sim   
 -e^{ -\fract{1}{M+1}\omega^{-(M{+}1)k} x^{M+1}}~,~~\nn\\[3pt]
y_k''  \sim   \omega^{-(M{+}1)k}x^{M} 
 e^{ -\fract{1}{M+1}\omega^{-(M{+}1)k} x^{M+1}}~,&&~
\label{largexk}
\eea
for 
$|x|\to\infty$ with
\eq
x\in\CS_{k-3/2}\cup\CS_{k-1/2}\cup\CS_{k+1/2}\cup\CS_{k+3/2}\,.
\label{sectr}
\en
Comparing 
$y_k$, $y_{k+1}$ and $y_{k+2}$ in the region
$\CS_{k+1/2}\cup\CS_{k+3/2}$, where the asymptotics of all three are
given by (\ref{largexk}),
establishes their linear independence. 
The  set $\{ y_k,y_{k+1}, y_{k+2}\}$ therefore
forms a basis of solutions 
to~(\ref{thirdotr}) (and, for $k$ integer, to
(\ref{thirdo})\,).
Alternatively, we can examine
\eq
W_{k_1,k_2,k_3}=W[y_{k_1},y_{k_2},y_{k_3}]\,,
\en
where 
the generalised Wronskian $W[f,g,h]$ is defined
to be
\eq
\mbox{Det}
\lf[\matrix{f & f' & f''  \cr
            g & g' & g''  \cr
            h & h' & h'' }\ri]\,.
\en
It is a standard result (see, for example, \cite{CL}\,)
that, for $f$, $g$ and $h$ solving
(\ref{thirdo}), $W[f,g,h]$ is independent of $x$, and that $f$, $g$
and $h$ are linearly independent if and only if $W[f,g,h]$ is 
nonzero.
For $(k_1,k_2,k_3)=(-1,0,1)$,  the 
asymptotic~(\ref{largexk}), used in 
$\CS_{-1/2}\cup\CS_{1/2}$, shows that
\eq
W_{-1,0,1}
= 8 i\sin (\fract{M}{3M+3 } \pi) \sin (\fract{2M}{3M+3 } \pi)~.
\label{W123}
\en
It is also the case that
\eq
W_{k_1+a,k_2+a,k_3+a}(E)= W_{k_1,k_2,k_3}(\omega^{-3Ma}E)~,
\label{wrel}
\en
so $W_{k,k+1,k+2}$ is nonzero for all $k$, thus confirming 
the independence of
 $\{ y_k,y_{k+1}, y_{k+2} \}$.

We now 
aim to generalise the analysis of \cite{DT4} to this situation,
guided in part by the treatment of $A_2$-related BA systems
provided by~\cite{PS}.
Since $y_1, y_2, y_3$ form a basis,  we can write
\eq
y_0- S^{(1)}(E) y_1 + S^{(2)}(E) y_2 - y_3=0
\label{tq1}
\en
with
\eq
S^{(1)}(E)= { W_{0,2,3} \over W_{1,2,3}}~~,~~~
S^{(2)}(E)= { W_{1,0,3} \over W_{1,2,3}}~.
\en
The coefficient of $y_3$ in (\ref{tq1}) 
is $-1$ by~(\ref{wrel}); $S^{(1)}$ and $S^{(2)}$ are Stokes
multipliers for (\ref{thirdo}), and are analytic functions of $E$.
Notice the formal similarity between this equation and eq.~(15) of
\cite{PS}.  

Now suppose that $k_1$ and $k_2$ differ by an integer. Then 
$y_{k_1}$ and
$y_{k_2}$ both solve (\ref{thirdotr})  
(with $e^{-2k\pi i}=
e^{-2k_1\pi i}=
e^{-2k_2\pi i}\,$), and
it can
be checked by direct substitution that the function
\eq
z_{k_1k_2}(x,E)=  y_{k_1}y'_{k_2}- y'_{k_1}y_{k_2}~
\label{zdef1}
\en
solves 
\eq
z'''_{k_1k_2}(x,E)
-
e^{-2k\pi i}
P(x,E) z_{k_1k_2}(x,E)=0\,.
\label{auxdif}
\en
This is just the equation adjoint to (\ref{thirdotr}); the 
observation that the Wronskian of two solutions of a third-order
ordinary differential equation satisfies
the adjoint equation dates back at least to Birkhoff~\cite{BIRK}\,.
{}Observe also that if $k_1$ and $k_2$ are shifted by a half-integer,
then a solution of the original equation (\ref{thirdotr}) results:
\eq
z'''_{k_1+\fract{1}{2}k_2+\fract{1}{2}}(x,E) + 
e^{-2k\pi i} P(x,E)
z_{k_1+\fract{1}{2}k_2+\fract{1}{2}}(x,E)=0~.
\label{half}
\en
For $|k_1-k_2|<3$, 
the regions (\ref{sectr}) for $k=k_1$ and $k=k_2$ have a nonempty
overlap, and
an asymptotic for $z_{k_1k_2}$ is easily
obtained from (\ref{largexk}).
In particular, for $k=1,2,3$ we have 
\eq
z_{-k/2,k/2}(x,E)\sim
2i\sin(\pi k/3)\,x^{-M}
e^{ -2\cos(\pi k/3)\fract{1}{M+1} x^{M+1}}~,\qquad x\to +\infty\,.
\label{eq0}
\en 
For $k=1$, $z_{-1/2,1/2}$ solves (\ref{thirdo}), and now
from (\ref{eq0}) we see that
it shares (up to a proportionality factor) the
asymptotic (\ref{largex}). By uniqueness, we deduce
\eq
z_{-1/2,1/2}(x,E) 
= i\sqrt{3}\, y(x,E)\,.
\label{zy}
\en
Unfortunately, this argument is not so effective for the other cases.
At $k=2$, the formula (\ref{eq0}) shows only that
$z_{-1,1}$ is not subdominant on the real axis, and this information
is not enough to pin the function down.
For $k=3$, $\sin(\pi k/3)=0$ and
all that can be deduced is
that the leading asymptotic of $z_{-3/2,3/2}$ is subleading to the
term over which we have control.

The next step is to manipulate 
(\ref{tq1}) in order to eliminate either $S^{(1)}$ or $S^{(2)}$. We
have
\bea
y'_1 y_0 - S^{(1)}(E) y'_1y_1 + S^{(2)}(E) 
y'_1 y_2 -  y'_1 y_3 &=&0\,;
\label{b1} \\
y_1 y'_0- S^{(1)}(E) y_1y'_1 
+ S^{(2)}(E) y_1y'_2 -  y_1y'_3&=&0\,,~~\label{b2}
\eea
and, subtracting,
\eq
S^{(2)}(E)z_{12}=z_{01}+z_{13}\,.
\label{szeq}
\en
For the reasons just explained, 
functions $z_{k_1k_2}$ with $|k_1-k_2|=1$ are the most easily handled,
so we use
the identity $y_2z_{13}=y_1z_{23}+z_{12}y_3$ to rewrite (\ref{szeq}) as
\eq
S^{(2)}(E) y_2 z_{12}
 = y_2 z_{01}+z_{12} y_3+z_{23} y_1\,.
\label{T2yy}
\en 
Likewise,
\eq
S^{(1)}(E)y_1 z_{12}=z_{12} y_0+ y_1 z_{23} + z_{01} y_2\,.
\label{T1yy}
\en
Now the result (\ref{zy}) can be combined with shifts in $E$ to
$\omega^{15M/4}E$ and $\omega^{21M/4}E$ respectively to
rewrite both
(\ref{T2yy}) and (\ref{T1yy}) as
\eq
T(E)y_{-{1/4}} y_{{1/4}} = y_{-{1/4}} y_{{5/4}}
+ y_{-{3/4}} y_{{3/4}}+ y_{-{5/4}} y_{{1/4}}~,
\label{tqqf} 
\en
where 
\eq
T(E)=
S^{(1)}(\omega^{15M/4}E)=
S^{(2)}(\omega^{21M/4}E)~.
\label{TSS}
\en
As a byproduct, this has
established that the two Stokes multipliers $S^{(1)}$ and
$S^{(2)}$ are related by an analytic continuation in $E$.

Finally, taking (\ref{tqqf}) at $x=0$ yields a functional relation
involving $E$ alone. To absorb various phases, it is convenient to
set
\eq
Q^{+}(E)=E^{-\frac{1}{3M}} 
y(0,E)~,~~
Q^{+}_k(E)=Q^{+}(\omega^{-3 M k}E)~.
\label{qy1}
\en
Then the relation is
\eq
T Q^{+}_{-{1/4}} Q^{+}_{{1/4}} = Q^{+}_{-{1/4}} 
Q^{+}_{{5/4}}
+ Q^{+}_{-{3/4}} Q^{+}_{{3/4}}+ Q^{+}_{-{5/4}} 
Q^{+}_{{1/4}}~.
\label{tqqf1} 
\en
This is very similar to the equations related
to the dilute $A$ model studied in~\cite{WNS,BNW}.
An equation involving
$y''(0,E)$ can also be derived. First, differentiate (\ref{szeq}) twice
with respect to $x$: 
\eq
S^{(2)}(E)z_{12}''=z_{01}''+z_{13}''\,.
\label{szeq1}
\en
Using the fact 
that $y_1$, $y_2$ and $y_3$ all solve (\ref{thirdo}), we have
$y_2'' z_{13}''=y_1'' z_{23}'' + z_{12}'' y_3''$, and so the previous
steps can be repeated to find
\eq
T(E)y_{-{1/4}}'' y_{{1/4}}'' = y_{-{1/4}}'' y_{{5/4}}''
+ y_{-{3/4}}'' y_{{3/4}}''+ y_{-{5/4}}'' y_{{1/4}}''~.
\label{tqqfa} 
\en
Again set $x=0$, and define
\eq
Q^{-}(E)=\fract{1}{2}E^{\frac{1}{3M}} 
y''(0,E)~,~~
Q^{-}_k(E)=Q^{-}(\omega^{-3 M k}E)~
\label{qy2}
\en
(the factor $\fract12$ is included for later convenience). Then
\eq
TQ^{-}_{-{1/4}} Q^{-}_{{1/4}} = Q^{-}_{-{1/4}} 
Q^{-}_{{5/4}}
+ Q^{-}_{-{3/4}} Q^{-}_{{3/4}}+ Q^{-}_{-{5/4}} 
Q^{-}_{{1/4}}~.
\label{tqqfa1} 
\en
There is no simple relation involving $y'(0,E)$ alone,  
but from~(\ref{zdef1}) and (\ref{zy}) one can deduce
\eq
i \sqrt{3} y'= y_{-1/2}y''_{1/2}- y''_{-1/2} y_{1/2} \,,
\label{yyy}
\en 
which  allows $y'(0,E)$ to be recovered once
$y(0,E)$  and  $y''(0,E)$ are known.
%
% 
%%%%%%%%%%%%%%%%%%%%%%%%%%%%%%%%%%%%%%%%%%%%%%%%%%%%
\resection{The non-linear integral equation}
\label{snie}
%%%%%%%%%%%%%%%%%%%%%%%%%%%%%%%%%%%%%%%%%%%%%%%%%%%%
%
%
The functions $Q^{\pm}(E)$ are not single-valued, and to derive an
integral equation it is more convenient to work with the
functions $y(0,E)$ and $y''(0,E)$ directly. Set
\eq
D^+(E)= y(0,E)~~,\quad D^-(E)=\fract{1}{2}y''(0,E)
\en
(so that $D^{\pm}(E)=E^{\pm\frac{1}{3M}}Q^{\pm}(E)$\, and 
$Q^{\pm}_k(E)=\omega^{\pm
k}E^{\mp\frac{1}{3M}}D^{\pm}(\omega^{-3Mk}E)$\,). These
are entire functions of $E$ and can be interpreted as
spectral determinants for the third-order equation (\ref{thirdo}),
since their zeroes coincide with the values of $E$ for which
the solution $y$, decaying at $x\to +\infty$ for
all values of $E$, in addition
either vanishes at $x=0$ (for the zeroes of $D^+$), or has
a vanishing second derivative at $x=0$ (for the zeroes of $D^-$).
(See, for example, ref.~\cite{DT4} for a more detailed discussion of this
point in the context of
second-order equations.)
For  $M>1/2$, the functions $D^{\pm}(E)$ have
large-$|E|$ asymptotics 
\eq
\ln D^{\pm}(E) \sim a_0 (-E)^{\mu}~~~~~|E| 
\rightarrow \infty~,~|\mbox{arg}(-E)|<\pi 
\label{bas}
\en
where  $\mu=(M+1)/3M$, $a_0 =\kappa(3M,3)$, and 
\eq
\kappa(a,b)=\int_0^{\infty}dx  
\lf ((x^{a}+1)^\fract{1}{b} -x^\fract{a}{b} \ri)
= { \Gamma(1+\fract{1}{a}) \Gamma(1+\fract{1}{b}) \over 
\Gamma(1+\fract{1}{a}+\fract{1}{b}) } 
{ \sin \fract{\pi}{b} \over  \sin (\fract{\pi}{b} +\fract{\pi}{a}) }~.
\en
The growth of $\ln D^{\pm}(E)$ is no
larger on the positive real
$E$-axis than elsewhere, so the orders of $D^+$ and $D^-$
as functions of $E$
are both equal to
$\mu$, and are less than $1$ for $M>1/2$. Invoking the Hadamard 
factorisation theorem, we can write
\eq 
D^{\pm}(E)= D^{\pm}(0) \prod_{k=1}^{\infty} 
\lf ( 1- {E \over E^{\pm}_k} \ri)\,. 
\label{pra}
\en
The  precise values of the  constants $D^{\pm}(0)$ are irrelevant 
for the treatment below, but some knowledge of the positions
of the  zeroes $\{E^{\pm}_k \}$ will be crucial.
We conjecture that, for all $M>0$, all of the zeroes of 
$D^{\pm}(E)$ lie on the  positive real $E$-axis. 
Some numerical evidence in favour of this claim will be presented 
below.

The generalised T-Q relations (\ref{tqqf1}), (\ref{tqqfa1}) taken at
either 
$E \in \{ \omega^{3M/4}E^{\pm}_n \}$ or
$E \in \{ \omega^{-3M/4}E^{\pm}_n \}$ imply
\eq
{D^{\pm}(\omega^{-3M}E_n^{\pm})\over D^{\pm}(\omega^{-3M/2}E_n^{\pm})} = 
-\omega^{\mp 1}
{D^{\pm}(\omega^{3M}E_n^{\pm})\over D^{\pm}(\omega^{3M/2}E_n^{\pm})}~,
\label{BAD}
\en
an equation that can be written in a Bethe-ansatz form as
\eq
\prod_{k=1}^{\infty}  { E^{\pm}_k - \omega^{-3M}    E^{\pm}_n \over  
E^{\pm}_k - \omega^{3M}  E^{\pm}_n}= - \omega^{\mp 1}
\prod_{k=1}^{\infty}  { E^{\pm}_k - \omega^{-3M/2} E^{\pm}_n \over  
E^{\pm}_k - \omega^{3M/2} E^{\pm}_n}~.
\label{BA}
\en
This equation is at least not inconsistent with the conjectured
reality of the $E_n$'s,
since both sides then reduce to pure phases.
There are certainly other, complex, solutions to (\ref{BA}),
so
the reality property should be seen as a way of selecting the
particular solution relevant to our differential equation, analogous
to the selection of the ground state in an integrable model.

A non-linear integral equation, similar to those 
described in~\cite{KBP,DDV,BLZ2}, can now be obtained for
the quantity
\eq
d^{\pm}(E)=  \omega^{\pm 1} 
 { D^{\pm}( \omega^{-3M} E  ) \over  D^{\pm}( \omega^{3M} E) }
 { D^{\pm}( \omega^{3M/2} E ) \over 
 D^{\pm}( \omega^{-3M/2} E) }~.
\en
We shall follow a path that completely parallels the treatment 
given in~\cite{BLZ2}. By (\ref{BAD}), $d^{\pm}(E)=-1$ at
the points $\{E^{\pm}_k\}$. (The value 
$-1$ might also occur at other points;
we supplement our previous conjecture with the assumption
that none of these points lie on the positive real
axis.)
The product representation~(\ref{pra}) implies
\eq
\ln d^{\pm}(E) = \pm i  \pi { 2 \over 3M+3}  +  \sum_{n=1}^{\infty} 
F(E/E_n)
\label{fe0}
\en
where
\eq
F(E) = \ln  {(1-E \omega^{-3M}) \over (1-E \omega^{3M}) }
{  (1-E \omega^{{3M/2}} )
\over (1-E \omega^{-{3M/2}}) } ~.
\en 
The sum over the $E_n$ in (\ref{fe0}) 
can be written as a contour integral 
\eq
\ln d^{\pm}(E) = \pm i  \pi { 2 \over 3M+3}+
 \int_C {d E' \over 2i \pi} F(E/E') \partial_{E'} \ln(1+d^{\pm}(E'))
\label{bhaskara}
\en
with the contour $C$ running from $+\infty$ to $0$ above the real axis,
winding around $0$ and returning to $+\infty$ below the real
axis. (It is at
this point that the conjectures about the locations of the $E_n$'s and
of the other zeroes of $d^{\pm}(E)+1$ are
used.)
If the new variable $\te=\mu\ln E$ is introduced, the function $F$
becomes
\eq
F(e^{3 M \te/(M+1)})= 
\ln \lf(\omega^{-\fr{3M}{2}} 
{ \sinh ( \fr{3}{2} \fr{1}{1+\xi} \te + i \pi \fr{\xi}{1+\xi}   )
  \over
  \sinh(    \fr{3}{2} \fr{1}{1+\xi} \te   -i \pi \fr{\xi}{1+\xi}  )       }
{ \sinh  (  \fr{3}{2} \fr{1}{1+\xi} \te + i \fr{\pi}{2}  \fr{1}{1+\xi}   ) 
  \over
  \sinh (  \fr{3}{2}  \fr{1}{1+\xi} \te - i \fr{\pi}{2}   \fr{1}{1+\xi}) 
     }\ri)~,
\en
with $\xi=1/M$.
Now define
\eq
f^{\pm}(\te)=\ln d^{\pm}(e^{3 M \te /(M+1)})~,
\en
use the property
$ d^{\pm}(E)^*= d^{\pm}(E^*)^{-1}$ and integrate by parts to recast
(\ref{bhaskara}) as
\bea
&&\ln f^{\pm}(\te ) -   \int_{-\infty}^{\infty} 
d\te'  R(\te-\te')   \ln f^{\pm}(\te'-i 0)
= \pm i   \pi { 2 \over 3M+3} \nn \\[3pt]
&&\qquad\qquad\qquad\qquad\qquad~~~{}-2 i  \int_{-\infty}^{\infty} 
d\te'  R(\te-\te') \Im m  \ln(1+f^{\pm}(\te'-i 0))
\qquad\qquad
\label{fe1}
\eea
with
\eq
R(\te)= { i \over 2 \pi  }  \partial_{\te} F(e^{3 M \te/(M+1)})~.
\en
The term $(1-R)*\ln f^{\pm}(\te)$ on the  LHS of~(\ref{fe1}) 
is  easily inverted using Fourier transforms. 
Using
\eq
i \partial_{\te} \ln 
{ \sinh( h \te + i \pi \tau) \over  \sinh( h \te - i \pi \tau) }
= {  2 h \sin (2 \tau \pi ) \over \cosh( 2 h \te)-  \cos(2 \tau \pi) }
\en
\eq
\int {d\te \over 2 \pi} e^{-i k \te}
 {2 h  \sin (2 \tau \pi)  \over \cosh( 2 h \te) -  \cos( 2 \tau \pi) }
= {\sinh( ( 1 -2 \tau  ) \fr{ \pi k}{2 h} ) 
\over  \sinh(\fr{\pi}{2}\fr{k}{h}) } ~,
 \en
we have 
\bea
&&{}~\tilde{R}(k)=
{\sinh(\fr{\pi}{3} (1 {-} \xi) k ) 
\over  \sinh(\fr{\pi}{3} (1{+}\xi) k) } 
+
{\sinh( \fr{\pi}{3} \xi k ) 
\over  \sinh(\fr{\pi}{3} (1{+}\xi) k)  }
={2\sinh(\fr{\pi k}{6})\cosh(\fr{\pi}{6}(1{-}2\xi)k)
\over \sinh(\fr{\pi}{3}(1{+}\xi)k)} ~,~~~~~\nn\\[4pt]
&&1-\tilde{R}(k)=
{\sinh(\fr{\pi}{3}\xi k)\cosh(\fr{\pi}{2}k)
\over\sinh(\fr{\pi}{3}(1{+}\xi)k)\cosh(\fr{\pi}{6}k)}~.
\eea
Transforming back to $\theta$ space and rewriting the imaginary part in
terms of values above and below the real axis, 
the  functions $f^{\pm}(\te)$ 
solve
\bea
f^{\pm}(\te)&=&
\pm i \pi \alpha  -i b_0 e^{\te}
+\int_{{\cal
C}_1}\!\varphi(\te{-}\te')\ln(1{+}e^{f^{\pm}(\te')})\,d\te'
\nn\\[2pt]
&&\qquad\qquad\qquad\qquad\qquad{}-
\int_{{\cal
C}_2}\!\varphi(\te{-}\te')\ln(1{+}e^{-f^{\pm}(\te')})\,d\te'
\label{nlie}
\eea
where $\alpha=2/3$,
the  contours ${\cal C}_1$
and ${\cal C}_2$ run from $-\infty$ to $+\infty$, just below and just
above the real $\te$-axis, 
\eq
\varphi(\te)=-\int_{-\infty}^{\infty}
\frac{e^{i k\te} \sinh( k \fr{\pi}{3}) \cosh( \fr{\pi}{6} k (1-2 \xi))} 
{\cosh( \fr{\pi}{2} k) \sinh( k \fr{\pi}{3} \xi )}%
\frac{d k}{2\pi}~~~,\qquad \xi=\fract{1}{M}\, ,
\label{krnl}
\en
and the constant $b_0=2 \sin(\pi \fr{M+1}{3M} ) a_0$ 
has been fixed using the asymptotic behaviour~(\ref{bas}).
(The corresponding zero mode can be traced to the zero in
$1-\tilde{R}$ at $k=i\,$.)
The parameter $\alpha$ in (\ref{nlie}) is analogous to the chemical
potential (or twist) term in the equations of~\cite{KBP,DDV,BLZ2}.

A first consistency check is immediate: in the large $\te$ limit 
of~(\ref{nlie}) the driving term $b_0 e^{\te}$ 
dominates, and so in this limit   the  
functions $\exp(f^{\pm}(\te))$ are  $-1$ at the points $\te=\te_n^{\pm}$,
or
\eq
E=E_n^{\pm} = e^{\te_n/\mu} = ((2n - 1 \pm \fr{2}{3}) \pi /b_0)^\fract{1}{\mu} 
{}~~,~~(n=1,2,\dots)~.
\label{scset}
\en
The same limit can be treated directly using a WKB-like
approach to the differential equation~(\ref{thirdo}). 
Start from~(\ref{secla}) with  $x>x_0$   and
fix $x_0$ to be at the  inversion point 
 $P(x_0,E)=0$ ($x_0=E^{1/3M}$).
Now, 
using analytic continuation  
(see for example~\S{47} in \cite{LaLi}),
the dominant part in the
region $x<x_0$ is
\eq
y(x,E) \sim 
|P(x,E)|^{-\fr{1}{3}}
\exp \lf(  \fr{1}{2} \int^{x_0}_{x}  |P(x,E)|^{\fr{1}{3}}  dx \ri)
 \cos\lf({\sqrt{3}\over 2}\int^{x_0}_x |P(x,E)|^{\fr{1}{3}}dx-\fr{\pi}{3}\ri)
\en
Thus to have $y(0,E)=0$, requires
\eq
\sqrt{3} \int^{x_0}_0 (x^{3M}-E_n^+)^{\fr{1}{3}}  = 
 b_0 (E^+_n)^{\mu} =(2 n-\fr{1}{3}) \pi~~,\qquad n=1,2,\dots\,, 
\label{wkbe}
\en
where the formula 
\eq
\int_{0}^{1} (1-x^{a})^\fract{1}{b} 
= {  \sin (\fract{\pi}{b} +\fract{\pi}{a}) \over
\sin \fract{\pi}{b}  } \kappa(a,b)~,
\en
was used.
The prediction~(\ref{wkbe}) agrees perfectly with~(\ref{scset}). 
In figure~1
the positions of the lowest
zeroes of  $D^{+}(E)$ are plotted in the range $ 0.1<3M<7$, and compared
with the WKB-like prediction. Evidence for the reality of the $E_n$ at
$3M=1$ will be given in the
next section; in the meantime, we note that the levels continue
smoothly away from that point, and the eigenvalues appear to remain real in
the range studied. (The figure can be compared with figures 1 and 2 of
\cite{DT4}, which illustrate cases where the spectrum does {\em not}
remain real in the full range displayed.)
\begin{figure}[h]
\vspace{-.5cm}
\centerline{
\resizebox{.75\linewidth}{.75\linewidth}%
{\includegraphics{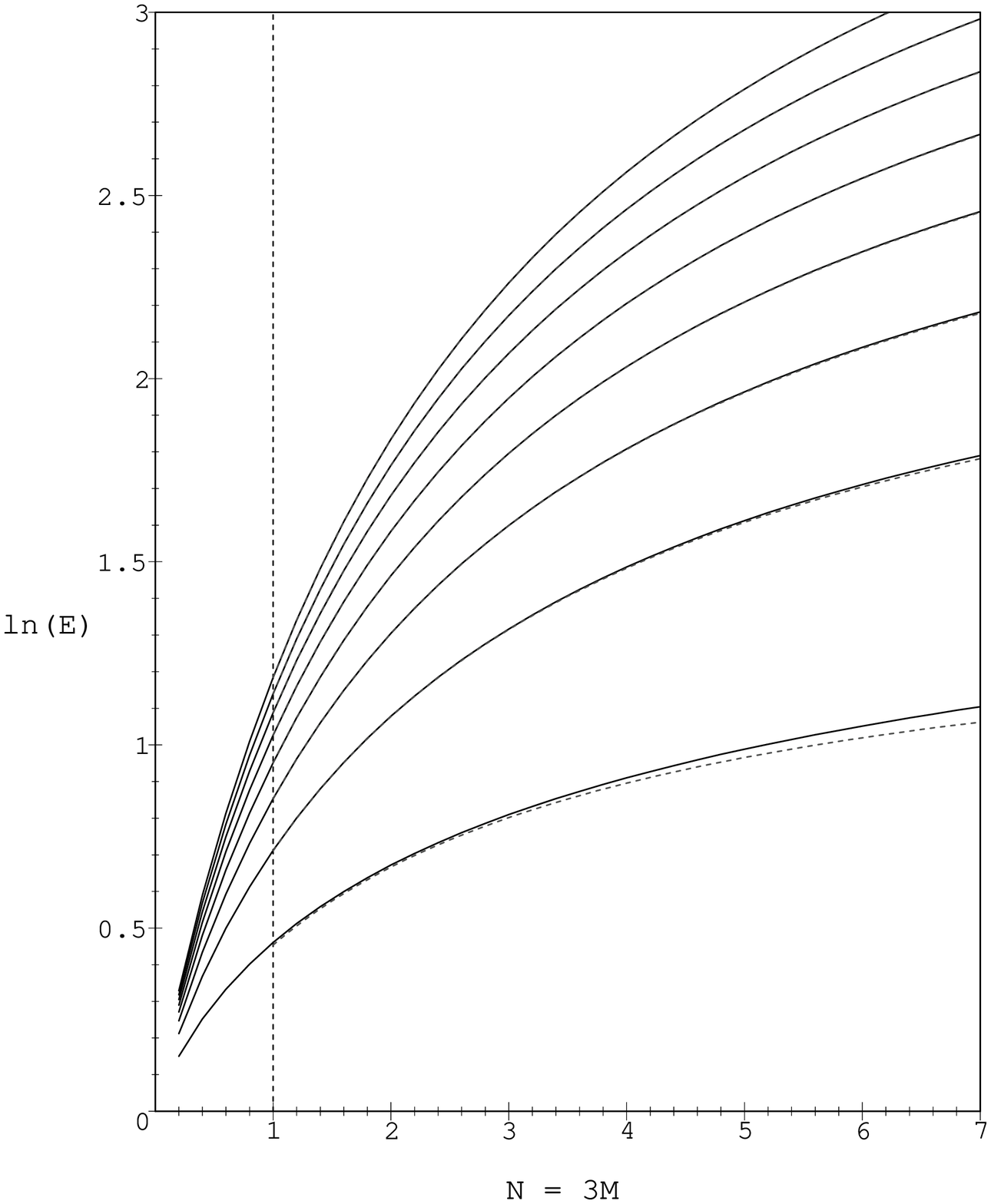}}
}
\centerline{
\parbox[t]{0.9\linewidth}%
{\small Figure 1: The positions of the first eight
zeroes
of $D^+(E)$, plotted on a log scale.
Dotted lines show the WKB-like predictions, and solid lines
the results from the nonlinear integral equation.}}
\label{figz}
\end{figure}
The kernel 
$\varphi(\te)$ given in~(\ref{krnl})  coincides
with  $i/2\pi$ times the logarithmic derivative of  the  scalar
factor in the Izergin-Korepin  S-matrix for the $a^{(2)}_2$
model \cite{IK} (cf.~eq.~(3.21) of \cite{Sm}, 
though note that the normalisation of $\xi$ used by Smirnov
in \cite{Sm} differs from ours: 
$\xi^{[{\rm Smirnov}]}
=\frac{2\pi}{3}\xi^{[{\rm this~paper}]}\,$.)
This is an element of the advertised
link between the differential equation
(\ref{thirdo}) and the $a^{(2)}_2$ model, the parameters being related
as $M=1/\xi$ (with $\xi$ related to the $a^{(2)}_2$ coupling $\gamma$
as $\xi=\gamma/(2\pi{-}\gamma)$~\cite{Sm}\,).
When  $3M$ is an  integer 
the potential is analytic,
and the associated scattering theory is diagonal;
the same phenomenon was observed in 
the Schr\"odinger/sine-Gordon case in~\cite{DT3}.  
The similarity 
between the relations
(\ref{tqqf1}), (\ref{tqqfa1}) 
and 
(\ref{BA}) 
and those arising in the
dilute $A$ model~\cite{WNS,BNW} has already been mentioned. Since the 
$a^{(2)}_2$ model is conjectured to be  
the continuum limit of the dilute $A$ model (see for example \cite{WPSN}), 
the fact that elements of it emerge  here is 
not a complete surprise. Nevertheless, it
is an encouraging signal that we are on the right track.
We will return to this point
in \S\ref{pcftsect}.
%
%%%%%%%%%%%%%%%%%%%%%%%%%%%%%%%%%%%%%%%%%%%%%%%%%%%%%%%%%%
%
% 
\resection{The linear potential}
\label{seclinpot}
A simple but non-trivial example occurs when $M=1/3$, and is the
analogue of the `Airy case' of the second-order problem, discussed in
\cite{DT3,Fendley}.
This lies outside the $M>1/2$ zone treated so far, so we
have to assume that the results obtained above continue to hold as the
region of their initial derivation is left.
The basic differential equation is
\eq
y'''(x,E)+xy(x,E)=Ey(x,E)~.
\en
Setting $y(x,E)=\CA(x-E)$, this becomes 
\eq
\CA'''(x) + x \CA(x)=0~.
\label{linpot}
\en
This equation
is solvable via a complex-Fourier transform:
\eq
\CA(x) =\sqrt{ 3 \over 2 \pi } 
\int_{\Gamma} e^{-i p x +{1 \over 4} p^4} dp~,
\label{gairy}
\en
where the integration   path $\Gamma$ is represented  in figure~2.
(A curious feature of this case is that the function $T(E)$ is a
constant, equal to $1$.)
\begin{figure}[h]
\vspace{-.2cm}
\centerline{
\resizebox{.4\linewidth}{.4\linewidth}%
{\includegraphics{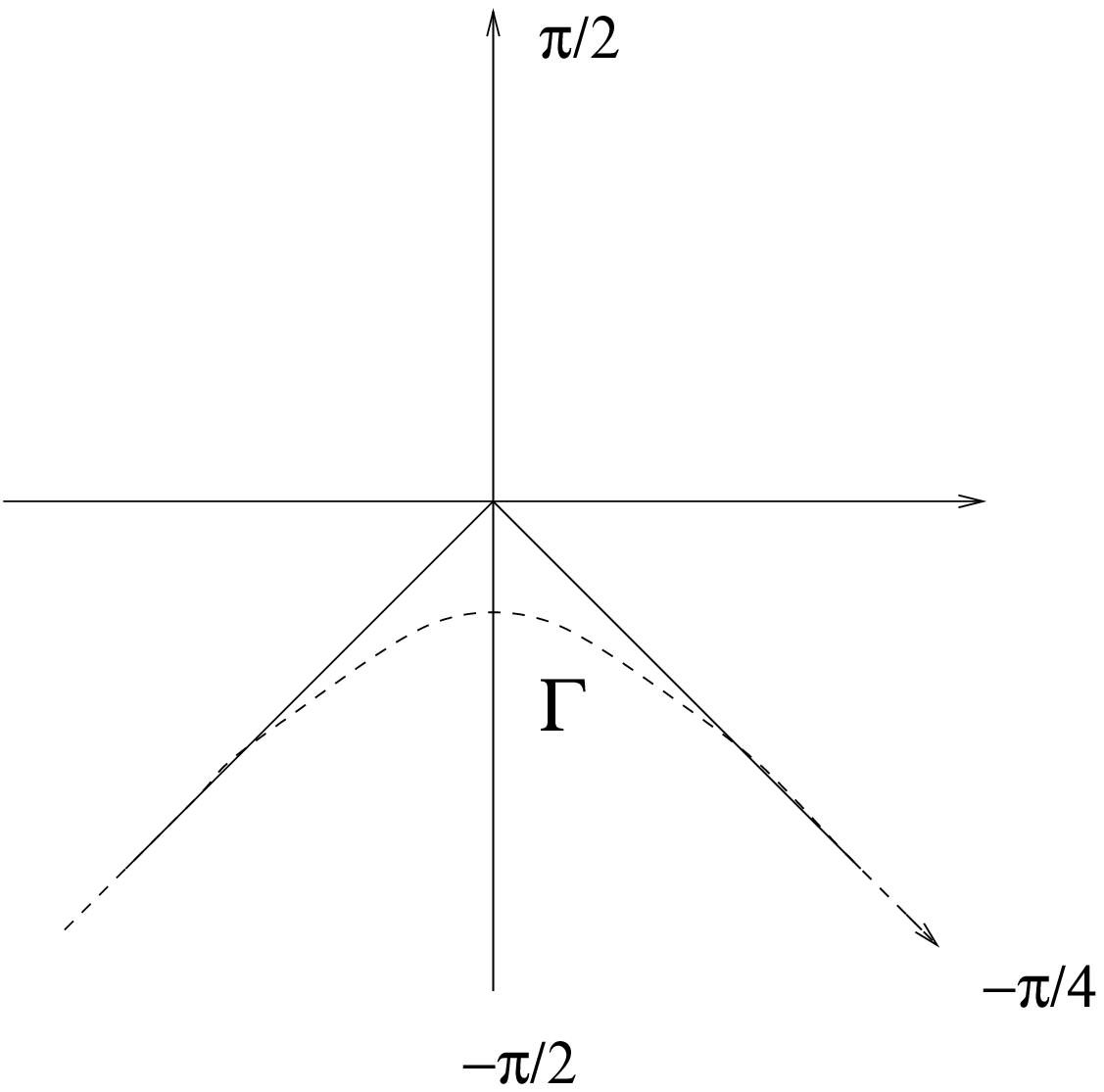}}
}
\noindent {\centerline{\small Figure 2: The integration path.~~~}}
\label{fig0}
\end{figure}
Even though the problem is not self-adjoint,
numerical evidence  suggests that 
all the zeroes  of  $\CA(x)$, $\CA'(x)$ and  $\CA''(x)$ lie on the
negative real axis  (see figure~3), and so the zeroes of 
$y(0,E) \equiv \CA(-E)$, and of $y'(0,E)$ and $y''(0,E)$,
 are  positive and real. 
In the first columns of tables~1 and 2,  the positions
of the first ten zeroes
of $\CA(-x)$  and  $\CA''(-x)$ are  displayed. 
\begin{figure}[h]
\vspace{-.01cm}
\centerline{
\resizebox{0.6\linewidth}{0.35\linewidth}%
{\includegraphics{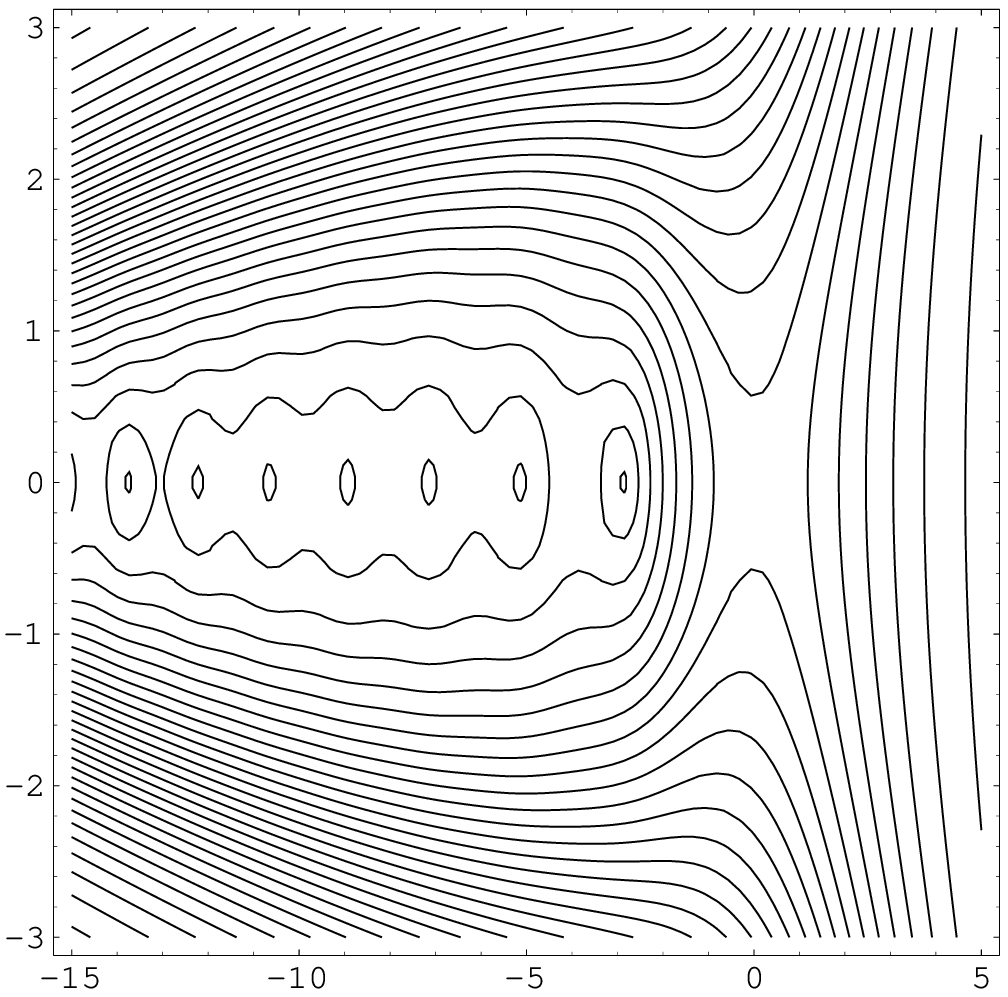}}}
\centerline{
\parbox[t]{0.7\linewidth}%
{\small Figure 3: A search for the zeroes of $\CA(x)$ in 
the complex $x$-plane. The function plotted is $|A(x)|/(1{+}|A(x)|)\,$, 
$A(x)= e^x \CA(x)\,$.}}
\label{fig3}
\end{figure}
%
%
%%%%%%%%%%%%%%%%%%%%%%%%%%%%%%%%%%%%%%%%%%%%%%%%%%%%%%%%%%
\begin{table}[htb]
\vspace{-0.2cm}
\begin{center}
\begin{tabular}{ c l l l  } 
\rule[-0.2cm]{0cm}{2mm} ~k~  &~~~~~$E_k$ (Exact)  &~~~~~$E_k$ (WKB)  
 &~~~~~$E_k$ (NLIE)
\\ \hline \rule[0.2cm]{0cm}{2mm} 
1\,    & \,2.8868281617697677  & \, 2.84467  & \, 2.886828161769766   \\
2\,    & \,5.1522519299627660  & \, 5.13866  & \, 5.152251929962763   \\
3\,    & \,7.1303732976716514  & \, 7.12265  & \, 7.130373297671650   \\
4\,    & \,8.9403621072563961  & \, 8.93513  & \, 8.940362107256395   \\
5\,    & \,10.635608688272157  & \, 10.6317  & \, 10.63560868827213   \\ 
6\,    & \,12.245164125544329  & \, 12.2421  & \, 12.24516412554437   \\
7\,    & \,13.787063381394688  & \, 13.7846  & \, 13.78706338139461   \\
8\,    & \,15.273489957153985  & \, 15.2714  & \, 15.27348995715393   \\
9\,    & \,16.713173447810789  & \, 16.7114  & \, 16.71317344781078   
\rule[-0.2cm]{0cm}{2mm} \\ 
\end{tabular}
\vspace{-0.1cm}
\caption{Zeroes  of $\CA(-x)$.} 
\end{center}
\end{table}
%%%%%%%%%%%%%%%%%%%%%%%%%%%%%%%%%%%%%%%%%%%%%%%%%%%%%%%%%%
%
%
%%%%%%%%%%%%%%%%%%%%%%%%%%%%%%%%%%%%%%%%%%%%%%%%%%%%%%%%%%
\begin{table}[htb]
\vspace{-0.2cm}
\begin{center}
\begin{tabular}{ c l l l  } 
\rule[-0.2cm]{0cm}{2mm} ~k~  &~~~~~$E_k$ (Exact)  &~~~~~$E_k$ (WKB)  
 &~~~~~$E_k$ (NLIE)
\\ \hline \rule[0.2cm]{0cm}{2mm}
1\,    & \, 0.8909448213691012   & \, 0.85075   & \, 0.890944821369104   \\
2\,    & \, 3.6439018898747455   & \, 3.66124   & \, 3.643901889874749   \\
3\,    & \, 5.8171363674931786   & \, 5.82455   & \, 5.817136367493172   \\
4\,    & \, 7.7377094704565507   & \, 7.74230   & \, 7.737709470456556   \\
5\,    & \, 9.5084853382296383   & \, 9.51174   & \, 9.508485338229636   \\ 
6\,    & \, 11.174539990553549   & \, 11.1770   & \, 11.17453999055357   \\
7\,    & \, 12.761111499098718   & \, 12.7631   & \, 12.76111149909874   \\
8\,    & \, 14.284208133811659   & \, 14.2859   & \, 14.28420813381167   \\
9\,    & \, 15.754819677279076   & \, 15.7562   & \, 15.75481967727909   
\rule[-0.2cm]{0cm}{2mm} \\                           
\end{tabular}
\vspace{-0.1cm}
\caption{Zeroes  of $\CA''(-x)$.} 
\end{center}
\vspace{-0.5cm}
\end{table}
%%%%%%%%%%%%%%%%%%%%%%%%%%%%%%%%%%%%%%%%%%%%%%%%%%%%%%%%%%
%

The approximate positions of the zeroes of $\CA(x)$ can be found from the
WKB formula of the last section. As a check, we
rederive them here via a saddle-point treatment of (\ref{gairy}).
The exponent of the integrand 
for $x=-|x|<0$ has  stationary points at
\eq
p_0=i |x|^\fract{1}{3}~~,~~ p_{\pm}= i e^{\pm i 2\pi/3}|x|^\fract{1}{3}~.
\en
Deforming  the contour $\Gamma$ so that it touches  the points $p_{\pm}$, we
get 
\eq
\CA(-|x|) \sim \CN (  
e^{ \fract{3}{4}|x|^{4/3} e^{-i  \pi/3} + i   \fract{\pi}{3}}    
+e^{\fract{3}{4} |x|^{4/3} e^{ i  \pi/3} - i \fract{\pi}{3} }  
)~,
\label{aairyas}
\en     
where the phases $\pm \pi/3$  are  the contributions   
from the choice of the steepest descent directions, 
transforming the quadratic terms in the expansion near the 
saddle points into a pure Gaussian integral  
\eq
\CN= \sqrt{ 3 \over 2 \pi } \int_{-\infty}^{\infty} 
e^{- \fract{3}{2} |x|^\fract{2}{3} t^2} dt
=  |x|^{-\fract{1}{3}}~.
\en  
Thus for large negative $x$ we have 
\eq
\CA(-|x|) \sim   2 |x|^{-\fract{1}{3}}   e^{\fract{3}{8} |x|^\fract{4}{3}} 
\cos( \fract{3}{8} \sqrt{3} |x|^\fract{4}{3} -   \fract{\pi}{3} )~.
\label{as1}
\en
For   $x> 0$, the dominant saddle point is instead at
$p_0=-i |x|^\fract{1}{3}$, 
and
\eq
\CA(x) \sim  |x|^{-\fract{1}{3}}  e^{- \fract{3}{4} |x|^\fract{4}{3}}~.
\en 
(This agrees with the general asymptotic (\ref{secla}),
since
$y(x,E)=\CA(x-E)$.)
The dominant behaviours of $\CA'(x)$ and $\CA''(x)$ 
for 
$x$ real and $|x|$ large are
\bea
\CA'(-|x|) \sim -2  e^{\fract{3}{8} |x|^\fract{4}{3}}
 \cos( \fract{3}{8} \sqrt{3} |x|^\fract{4}{3})~~&,&~~
\CA'( |x|) \sim - e^{- \fract{3}{4} |x|^\fract{4}{3}}~~~,  \\
\CA''(-|x|) \sim 2 |x|^{\fr{1}{3}} e^{\fract{3}{8} |x|^\fract{4}{3}}
 \cos( \fract{3}{8} \sqrt{3} |x|^\fract{4}{3}+ \fract{\pi}{3})~~&,&~~
\CA''( |x|) \sim |x|^{\fr{1}{3}} e^{- \fract{3}{4} |x|^\fract{4}{3}}~. 
\label{as3}
\eea
Continuing to differentiate, the general result for the approximate
positions of the zeroes of the $m^{\rm th}$ derivative
$\CA^{(m)}(x)$ is
\eq
\CA^{(m)}(x)=0~~:~ 
x =- \lf({4 \over 3 \sqrt{3}} (2n- {2m+1 \over 3}) \pi \ri)^{3/4}~~
{}~~(n=1,2,\dots)~~. 
\label{zeroas0}
\en
At  $m=0$ (\ref{zeroas0}) reduces to the $M=1/3$ WKB prediction~(\ref{wkbe}).
In tables 1 and 2 the results 
from  formula~(\ref{zeroas0})  are compared with the `exact' result 
from a numerical treatment 
of~(\ref{gairy}), and also against the results of the numerical
solution of the nonlinear integral equation (\ref{nlie}). Clearly the
agreement is very good.
\resection{Duality and a more general 
chemical potential}
\label{dualitysec}
In this section  we shall  
investigate the effect of a  duality transformation
on~(\ref{thirdo}), analogous to that studied in~\cite{BLZschr}
for the Schr\"odinger equation~(\ref{Sch}).
In the Schr\"odinger case   duality maps wavefunctions for   
confining potentials ($M>0$)   to 
wavefunctions for singular potentials  ($-1<M<0$),
in such a way that the theories with $M$ and  $\tilde{M}=-M/(M+1)$
are dual, their respective spectral problems being essentially
equivalent. It also changes the coefficient of the `angular momentum'
term $l(l{+}1)/x^2$ in (\ref{Sch})
in a non-trivial way; the same phenomenon 
here will allow us to guess the corresponding
term for the third-order problem (\ref{thirdo}).

To implement the duality transformation, we begin with a
Langer\,\cite{langer} -type  variable transformation
\eq
y(x)= e^{z} u(z)~~,~~z=\ln x~,
\label{dualtra}
\en
after which~(\ref{thirdo}) becomes
\eq
u'''(z)-u'(z)+ (e^{(3M+3)z} - E e^{3 z }) u(z)=0~.
\label{morse1}
\en
The duality   $M \to \tilde{M}$ is now effected
by interchanging
the r\^oles of the two exponentials.  Substituting 
$ z \rightarrow \fract{z}{M+1}+ \ln \fract{M+1}{E^{1/3}} $
yields 
\eq
u'''(z) -{1 \over (M+1)^2} u'(z) +(-e^{3 z/(M+1)} - \tilde{E} e^{3 z}) u(z)=0~,
\label{morse2}
\en
where $\tilde{E}=-(M{+}1)^{3M}/E^{M+1}$.

Now transforming back results in the equation
\eq
\tilde{y}'''+ {M (M{+}2) \over (M{+}1)^2} 
(\frac{1}{x^2}\frac{d}{dx}-\frac{1}{x^3})\tilde{y}
+(-x^{-\fract{3M}{M+1}}  -\tilde{E}) \tilde{y}=0~.
\label{dual0}
\en
As promised, the confining `potential' $x^{3M}$ has been exchanged for
a singular potential $-x^{-3M/(M{+}1)}$, and a new term, proportional
to $(x^{-2}d/dx-x^{-3})y$, has been generated.
This  motivates us to enlarge the set of differential equations under
consideration to
\eq
y'''-G(\frac{1}{x^2}\frac{d}{dx}-\frac{1}{x^3})y
+(x^{3M}-E)y=0
\label{enlarged}
\en
with $G$ a new parameter, analogous to $l(l{+}1)$ for the Schr\"odinger
equation.
Duality maps (\ref{enlarged}) to
\eq
\tilde{y}'''- \tilde{G}
(\frac{1}{x^2} \frac{d}{dx}-\frac{1}{x^3})\tilde{y}+
(-x^{3\tilde{M}}-\tilde{E})\tilde{y}=0~,
\label{endual}
\en
where
\eq
\tilde{y}(x,\tilde{E},\tilde{G})=
(M{+}1)^{-1} E^{1/3}
 x^{M \over (M+1)} 
y ((M{+}1)E^{-1/3} 
x^{1/(M{+}1)}, E,G)
\label{ytildey}
\en
and
\eq
\tilde{M} =- {M \over M+1}
{}~~,~~~
\tilde{E}= -\frac{(M{+}1)^{3 M}}{E^{M+1}}
{}~~,~~~
\tilde{G}={G-M (M{+}2) \over (M{+}1)^2}~.
\label{dualpar}
\en
Duality therefore
maps the 3-parameter family  ($\{M,E,G\}$) 
of differential equations~(\ref{enlarged}) onto itself.
The analysis of \S2 can now be repeated for these generalised
problems. It is convenient to write $G=g(g{+}2)$ and to work mostly
with $g$ instead of $G$.
We first enlarge the scope of (\ref{ykdef}) by setting
\eq 
y_k(x,E,g)= \omega^k y(\omega^{-k}x,\omega^{-3M k}E,g)~; 
\en
then $y_k$ solves
\eq
y_k'''  
- G (\frac{1}{x^2}\frac{d}{dx}-\frac{1}{x^3}) y_k
+ e^{-2 k \pi i }P(x,E) y_k 
=0~. 
\label{gendifa}
\en
Next we define
$z_{k_1 k_2}$ as in~(\ref{zdef1}). If 
$e^{-2 k_1 \pi i}=e^{-2 k_2 \pi i}\equiv e^{-2 k \pi i}$,
then 
\eq
z'''_{k_1 k_2}  
- G (\frac{1}{x^2}\frac{d}{dx}-\frac{1}{x^3}) z_{k_1 k_2} 
- e^{-2 k \pi i }P(x,E)
 z_{k_1 k_2}=0~, 
\label{auxdif1}
\en
which is the equation adjoint to 
(\ref{gendifa})\footnote{Notice 
that the operator $x^{-2}d/dx-x^{-3}$ is by itself anti-self-adjoint;
this gives some insight as to why it is a sensible
generalisation of the $x^{-2}$ term in the Schr\"odinger equation.}.
We can recover the original problem
by shifting $k$ by a half-integer, and arguing just as before we find
that
\eq
z_{-1/2,1/2}(x,E,g)=   
i\sqrt{3}\, y(x,E,g)~
\en 
and then
\eq 
T(E,g) y_{-1/4}\,y_{1/4}= 
 y_{-1/4}\,  y_{5/4}+ 
y_{-3/4}\, y_{3/4}+ 
y_{-5/4}\, y_{1/4}~,
\label{gensa}
\en 
where
$T(E,g)=
S^{(1)}(\omega^{15M/4}E,g)=
S^{(2)}(\omega^{21M/4}E,g)\,$,
and $S^{(1)}$ and $S^{(2)}$ are defined as in (\ref{tq1}).
A non-zero value of $G=g(g{+}2)$ 
causes (\ref{gendifa}) to be singular at the
origin, so simply considering (\ref{gensa}) at $x=0$ is not an
option. Instead,
we expand $y(x,E,g)$ as
\eq
y(x,E,g) =  D^{+}(E,g)~\chi_+ + 
D^{0}(E,g)~\chi_0 +   D^{-}(E,g)~\chi_{-}~, 
\label{specdet2}
\en
where  $\{ \chi_+, \chi_0, \chi_- \}$
forms a basis of solutions defined via behaviour near the origin:
\eq
\chi_i(x,E,g)  \sim   x^{\lambda_i}+O(x^{\lambda_i+3})~, \qquad i=+,0,-\,,
\en
with the $\lambda_i$'s the roots of 
the indicial equation 
$(\lambda-1)(\lambda (\lambda-2)-g(g+2))=0\,$:
\eq
\lambda_{+}  =  -g~,~~~
 \lambda_{0} = 1 ~,~~~
\lambda_{-}  = g+2  ~.
\label{lg}
\en
Explicitly, the functions $D^+$,  $D^0$ and $D^-$ are
\eq
D^+=\frac{W[y,\chi_0,\chi_-]}{W[\chi_+,\chi_0,\chi_-]}~~,\quad
D^0=\frac{W[y,\chi_-,\chi_+]}{W[\chi_+,\chi_0,\chi_-]}~~,\quad
D^-=\frac{W[y,\chi_+,\chi_0]}{W[\chi_+,\chi_0,\chi_-]}~~,~~~
\en
with $W[\chi_+,\chi_0,\chi_-]=2\,(g{+}1)^3$. At $g=0$ they
reduce to $y(0,E)$, $y'(0,E)$ and $y''(0,E)/2$ respectively, in
agreement with the notation of earlier sections.

Defining
\eq
Q^{\pm}(E,g)=E^{\mp \fract{g+1}{3M}} 
D^{\pm}(E,g)~,~~~~
Q^{\pm}_k =Q^{\pm}(\omega^{-3 M k}E,g)~,
\label{QDdef2}
\en
the generalised T-Q relations (\ref{tqqf1}), (\ref{tqqfa1}) have exactly
the same form as before,
and (\ref{BA}) becomes 
\eq
\prod_{k=1}^{\infty}  { E^{\pm}_k - \omega^{-3M}    E^{\pm}_n \over  
E^{\pm}_k - \omega^{3M}  E^{\pm}_n}= - \omega^{\mp (g{+}1)}
\prod_{k=1}^{\infty}  { E^{\pm}_k - \omega^{-3M/2} E^{\pm}_n \over  
E^{\pm}_k - \omega^{3M/2} E^{\pm}_n}~.
\label{BA2}
\en
The arguments of \S\ref{snie} can now be repeated essentially verbatim, to
discover that, so long as $G$ is such that the conjectures of \S3 about the
zeroes of $T$ and $D^{\pm}$ remain true,
the quantities $D^+$ and $D^-$ for the more general
differential equation (\ref{enlarged})
are again described by the nonlinear integral
equation (\ref{nlie}), but 
with chemical potential term now taking the value 
 $\alpha=\fract{2}{3}\,(g{+}1)\,$.

At $M=1$,  $\omega^{\pm 3M}=-1$, the  LHS of 
(\ref{BA2}) is  $1$ and the $A_2$-related BA equation `collapses'
onto one more closely linked with $A_1$.
This leads to a rather surprising 
equivalence between 
spectral problems
for a  Schr\"odinger equation with
potential $x^6 + l(l+1)/x^2$~and the third-order problem~(\ref{enlarged}) 
at  $M=1$.
For these special points,
the quantities in this paper are related to those
of~\cite{DT4} as
\eq
D^{\pm}(E,g)|_{M=1}\propto
D^{\pm}(c^{-3/2}E,l)|_{M=3}^{\rm [ref.\hbox{\small\cite{DT4}}]}~~,
\quad
T(E,g)|_{M=1}=
T_{1}(c^{-3/2}E ,l)|_{M=3}^{\rm [ref.\hbox{\small\cite{DT4}}]}~,
\en
where $l{+}\frac{1}{2}=\frac{2}{3}\,(g{+}1)\,$, $c=
\frac{27}{4} {\Gamma(7/6) \over  \sqrt{\pi}}
{\Gamma(5/3) \over  \Gamma(1/3)}\,$, and $T_1$ is one of the 
`fused' $T$-operators discussed in \S4 of \cite{DT4}.
%
%%%%%%%%%%%%%%%%%%%%%%%%%%%%%%%%%%%%%%%%%%%%%%%%%%%%%%%%%%%%%
%
\resection{A link with perturbed conformal field theory}
\label{pcftsect}
We now return to the relation with the $a^{(2)}_2$ model, briefly
mentioned at the end of \S3.
The analogy with  results of~\cite{KBP,DDV} for the $A_1$-related models
suggests that the quantity
\eq
c_{\rm eff}= {6 i b_0 \over \pi^2} 
 \lf(\int_{{\cal
C}_1} e^{\te} \ln(1{+}e^{f^{\pm}(\te)})\,d\te-
\int_{{\cal
C}_2} e^{\te}\ln(1{+}e^{-f^{\pm}(\te)})\,d\te \ri)~
\label{itc}
\en
should be interpreted
as an effective central charge of an underlying conformal field
theory. 
For general $\alpha=\fract{2}{3}\,(g{+}1)\,$, this would predict
\eq
c_{\rm eff}=1- {3 \over M+1} \alpha^2 \, .
\label{ceff1}
\en 
Going further,
it is natural to interpret (\ref{nlie}) and (\ref{itc}) as the 
ultraviolet limit
of the following `massive' system:
\[f(\te)=
 i \pi \alpha  -i r\sinh\theta
+\!\int_{{\cal
C}_1}\!\!\varphi(\te{-}\te')\ln(1{+}e^{f(\te')})\,d\te'
-\!\int_{{\cal
C}_2}\!\!\varphi(\te{-}\te')\ln(1{+}e^{-f(\te')})\,d\te'\,;
\]
\eq
c_{\rm eff}(r)= {3 i r \over \pi^2} 
 \lf(\int_{{\cal
C}_1} \sinh\theta\, \ln(1{+}e^{f(\te)})\,d\te-
\int_{{\cal
C}_2}\sinh\theta\,\ln(1{+}e^{-f(\te)})\,d\te \ri)\,.
\label{mnlie}
\en
This should encode finite-size
effects in the massive $a_2^{(2)}$ theory, with $r=M_sR$,
$M_s$ the mass of the fundamental soliton
and 
$R$   the circumference of the (infinite) 
cylinder on which  the theory is living.
(Notice that there is no need to distinguish $f^+$ from 
$f^-$ any more, 
since the mapping $\theta\to -\theta$ now has the effect
of negating $\alpha$.)
There is now a natural scale,  which can be related to
an operator $\phi$ perturbing the ultraviolet conformal field theory. 
Standard  considerations~\cite{Zam2},
based on the $\te\to\te +i 2 \pi \fract{M+1}{3M} $  periodicity of 
$f(\te)$, suggest 
that so long as $\alpha$ is not an integer $c_{\rm eff}(r)$ will have 
an expansion in powers of $r^{6M/(M{+}1)}$ (together with an irregular 
`anti-bulk' term, irrelevant to the current discussion). This implies 
for $\phi$ either the conformal dimensions
\eq
\bar{h}_{\phi}=h_{\phi}= 1-{3M \over 2M{+}2}
\label{dim}
\en 
and an expansion of $c_{\rm eff}(r)$ in which only even powers of 
the coupling $\lambda$ to the operator $\phi$ appear, 
or, alternatively, the conformal dimensions
\eq
\bar{h}_{\phi}=h_{\phi}= 1-{3 M \over  M{+}1}
\label{dim1}
\en
and an expansion which sees both even and odd powers of $\lambda$.

When $\alpha$ is an integer, the standard considerations of \cite{Zam2}
may have to
be modified. Absorbing the term $i\pi\alpha$ into a shift in 
$f(\theta)$,
(\ref{mnlie}) becomes exactly
odd under a negation of $\theta$. This forces the 
shifted $f(\theta)$ to be zero 
at $\theta=0$, even in the far ultraviolet, and so long as the
would-be plateau value is nonzero, it splits
the plateau region into two
pieces, each of half the previous length.
As a result, the regular expansion of $c_{\rm eff}(r)$ 
is in powers of $r^{3M/(M{+}1)}$, and not $r^{6M/(M{+}1)}$.
Formula (\ref{dim}) now describes the situation
when both even 
and odd powers of $\lambda$ appear in 
the expansion of $c_{\rm eff}$, 
while for even powers only, the correct formula is
$\bar h_{\phi}=h_{\phi}=1-3M/(4M{+}4)$.
Note though that
this plateau-splitting effect does not occur at $\alpha=0$, since for
this case the plateau value of $f$ is anyway zero, and imposing
$f(0)=0$ has no effect.

As explained 
in~\cite{Sm} (see also \cite{KMM,koubek,Ef}), 
the  $a^{(2)}_2$ model, when 
appropriately quantum-reduced,
should correspond to the minimal models ${\cal M}_{p,q}$ (with $p$ and $q$ 
coprime integers 
and $p<q$)
perturbed by either
$\phi_{12}$, $\phi_{21}$ or \cite{MJM,GT} $\phi_{15}$. 
 With $\phi_{12}$ the perturbing operator,
the relation with the
parameter $\xi$ appearing in the kernel (\ref{krnl}) is~\cite{Sm}
\eq
\frac{p}{q}=\frac{2\xi}{(1{+}\xi)}~.
\label{pqrel}
\en
Since $M=1/\xi$, the ultraviolet
effective central charge
for a given value of $\alpha$, as predicted by (\ref{ceff1}), is
\eq
c_{\rm eff}=1-{3 p \over 2 q} \alpha^2 \, .
\label{ceff2}
\en 
To recover $\phi_{21}$ perturbations one simply has
to swap $p$ and $q$ in (\ref{pqrel}) and (\ref{ceff2})~\cite{Sm},
while
to find  $\phi_{15}$,
$p/q$ should be replaced by $4p/q$~\cite{MJM}.
  
For the  
sine-Gordon model,  naturally associated with the 
$\phi_{13}$ perturbing operator~\cite{Sms,lecl},
it has been observed both analytically and 
numerically~\cite{KBP,DDV,BLZ2,FMQR,FRT3}
that reduction is implemented
at the level of 
finite-size effects and 
the non-linear integral equation 
via a particular choice of the chemical potential.
The similarity between  our equations  and those 
in~\cite{KBP,DDV} 
suggests
that the same should be true here. To decide which
value of $\alpha$ will
tune~(\ref{mnlie})  onto  the ground state
of the relevant perturbed minimal model, we demand that the ultraviolet
effective central charges match up; 
the predicted dimensions of the perturbing operators,
and a comparison of results at
nonzero values of $r$ with those obtained via the thermodynamic Bethe
ansatz method, will then provide some nontrivial tests of the proposal.
 
The effective central charge 
of the ground state 
of the theory ${\cal M}_{pq}$
is  $c_{\rm eff}=1-6/pq$. 
Thus to have any chance of matching the vacua  of the
 $\phi_{12}$-perturbed  models,
we must set $\alpha=2/p$.
The required value of $h_{\phi}$, namely $h_{12}=\frac{3p}{4q}-
\frac{1}{2}$, is then 
matched by (\ref{dim}). 
The value just chosen for
$\alpha$ being a non-zero integer if and only if $p=2$, 
(\ref{dim}) will be the correct formula to use
provided the regular parts of the ground state energies of the 
models ${\cal M}_{pq}$ perturbed by $\phi_{12}$ expand in even powers of
$\lambda$ for $p\ge 3$, and in even and odd powers for $p=2$. 
This `prediction' holds for all of the examples that we checked.
We then compared numerical
results for $(M,\alpha)=(4,1)$, $(\fract{5}{3},\fract{2}{3})$, 
$(\fract{3}{2},\fract{1}{2})$,
 $(\fract{4}{3},\fract{1}{3})$ and $(1,0)$
against the tables of~\cite{Zam1,KM} 
for the $A_2^{(2)}$ (Yang-Lee), $E_8$, $E_7$, $E_6$ and $D_4$-related 
TBA equations, respectively, finding excellent agreement.
Swapping $p$ and $q$ in (\ref{ceff2}), 
the choice $\alpha=2/q$  should capture the $\phi_{21}$ cases.
The conformal weight of $h_{21}=\frac{3q}{4p}-
\frac{1}{2}$ is matched by (\ref{dim}), provided the swap of $p$
and $q$ in (\ref{pqrel}) is remembered. This time
$\alpha$ is never an integer, and the use of (\ref{dim})
is justified by the
regular parts of the ground state energies of the
$\phi_{21}$ perturbations always being in even powers of the
coupling $\lambda$.
For $(M,\alpha)=(\fract{1}{2},\fract{1}{2})$, 
$(\fract{2}{3},\fract{1}{3})$ and $(\fract{1}{5},\fract{2}{5} )$ 
the results from the $A_1$ and
$A_2$-related  TBA equations~\cite{Zam1} and the ${\cal{M}}_{35}$ 
model~\cite{RST,BP} were reproduced within our numerical 
accuracy\footnote{Beware of a misprint in eq.~(7) of \cite{RST}:
the (minus) sign before   $\sum_{j=1}^2$ 
should be reversed.}. Finally, replacing 
$p/q$ by $4p/q$
in (\ref{ceff1})
($ q>2p$), at $\alpha=1/p$ the models 
${\cal{M}}_{pq}$ perturbed by $\phi_{15}$
are recovered. This time it is (\ref{dim1}) which 
predicts the correct value for $h_{\phi}$,
as expected given that $\phi_{15}$ 
perturbations expand in both even and odd powers of $\lambda$. 
TBA equations for a number of  $\phi_{15}$-perturbed
models have been proposed
in~\cite{MJM,RST,KTW}, but so far we have
only compared (\ref{nlie}) with the TBA for the $\phi_{15}$ perturbation
of the ${\cal{M}}_{37}$ model given in~\cite{RST}.   
A selection  of our numerical results for all of the cases
just mentioned is presented in table~3, 
together with thermodynamic Bethe ansatz 
data taken from refs.~\cite{Zam1,KM,RST,BP}.
To facilitate the comparision,
we took $r=M_1R$ throughout, with $M_1$ 
the mass of the fundamental particle in the reduced
scattering theory. For 
$A_2^{(2)}$ and $E_8$ 
this is the first breather in the unreduced theory, and
$M_1$ is equal to $2 \cos( 5 \pi/12) M_s$ and $ 2 \cos(3 \pi/10)M_s$ 
respectively.
In all of the other models in the table, $M_1$ is equal to $M_s$.
The results strongly support the claim
that the system
(\ref{mnlie}) 
encodes the ground state energies of $\phi_{12}$, $\phi_{21}$ and
$\phi_{15}$
perturbations of minimal models. 
%
%%%%%%%%%%%%%%%%%%%%%%%%%%%%%%%%%%%%%%%%%%%%%%%%%%%%%%%%%%
\begin{table}[htb]
\begin{center}
\vskip 3mm
\begin{tabular}{c  c | l l l  } 
Model & \rule[-0.2cm]{0cm}{2mm} ~$(M,\alpha)$~&~~~$r$~~~&~~~~~TBA  
~ &~~~~~NLIE
\\ \hline 
\rule[0.2cm]{0cm}{2mm} 
$A_2^{(2)}
+\phi_{12}
$ & ($4,1$)     
     & 0.001  &  0.399999735051974  & 0.399999735051971  \\
&    & 0.002  &  0.399998953903823  & 0.399998953903824  \\ \cline{1-5}
\rule[0.2cm]{0cm}{2mm} 
$E_8
+\phi_{12}
$ &($\fr{5}{3},\fr{2}{3}$) 
     & 0.025  &  0.499926331494289  & 0.499926331494288  \\
&    & 0.05   &  0.499705463734389  & 0.499705463734387  \\ \cline{1-5}
\rule[0.2cm]{0cm}{2mm} 
$E_7
+\phi_{12}
$ &($\fr{3}{2},\fr{1}{2}$) 
     & 0.02   &  0.699928050129612  & 0.699928050129611  \\
&    & 0.04   &  0.699712371203531  & 0.699712371203531  \\ \cline{1-5}
\rule[0.2cm]{0cm}{2mm} 
$E_6
+\phi_{12}
$ &($\fr{4}{3},\fr{1}{3}$) 
     & 0.025  &  0.857016839032789  & 0.857016839032790  \\
&    & 0.05   &  0.856639509661813  & 0.856639509661819  \\ \cline{1-5}
\rule[0.2cm]{0cm}{2mm} 
$D_4
+\phi_{12}
$ &($1,0$)     
     & 0.01   &  0.999972507850553  & 0.999972507850552  \\ 
&    & 0.02   &  0.999890328583463  & 0.999890328583464  \\ \cline{1-5}
\rule[0.2cm]{0cm}{2mm} 
$A_1
+\phi_{21}
$ &($\fr{1}{2},\fr{1}{2}$) 
     & 0.02   &  0.499697279140833  & 0.499697279140832  \\
&    & 0.04   &  0.498957654198721  & 0.498957654198722 \\ \cline{1-5}
\rule[0.2cm]{0cm}{2mm} 
$A_2
+\phi_{21}
$ & ($\fr{2}{3},\fr{1}{3}$) 
     & 0.001  & 0.799999470103948    & 0.799999470103940  \\
&    & 0.002  & 0.799997907807646    & 0.799997907807649  \\ \cline{1-5}
\rule[0.2cm]{0cm}{2mm} 
$M_{35}
+\phi_{21}
$ & ($\fr{1}{5},\fr{2}{5}$) 
     & 0.1    &  0.596517064916761   & 0.596517064916762  \\
&    & 0.15   &  0.592881408017592   & 0.592881408017593  \\ \cline{1-5}
\rule[0.2cm]{0cm}{2mm} 
$M_{37}
+\phi_{15}
$ & ($\fr{1}{6},\fr{1}{3}$) 
     & 0.1   &   0.709591770021299   & 0.709591770021299 \\
&    & 0.15  &   0.705031895238354  & 0.705031895238357  \\ \cline{1-5}
\rule[-0.2cm]{0cm}{2mm} \\ 
\end{tabular}
\caption{NLIE results versus TBA data from 
refs.{\protect~\cite{Zam1,KM,RST,BP}}}
\end{center}
\vspace{-0.2cm}
\end{table}

In previously-studied examples, equations similar in form 
to those for the ground state have been found to describe 
excited states (see for example~\cite{BLZ2,BLZ4,DT1,FMQR}). 
We expect that the same will be possible here,
but we will leave
investigation of this point for future work.
Finally,  we remark that it would be interesting to derive
a nonlinear integral equation
for the $a^{(2)}_2$  model directly from
finite lattice  BA equations. 
We understand that progress is currently being made in this direction
\cite{BFR}.
\resection{General  $A_2$-related BA equations }
\label{a2sec}
In this section, we discuss the effect  
of adding a term proportional to $x^{-3}$ to
the differential equation~(\ref{enlarged}). 
The equation becomes
\eq
y'''-G(\frac{1}{x^2}\frac{d}{dx}-\frac{1}{x^3})y
+P(x,E,L)y =0~~,
\label{finaldif}
\en
where
\eq
P(x,E,L)= x^{3M} - E +{L \over x^3}~.
\en
Duality acts on $M$ and $G$ as before, and transforms   $L$ as
\eq 
L \to \tilde{L}= {L \over (M+1)^3}~.
\en
The relation~(\ref{ytildey}), 
apart from the appearance  of  $L$ and $\tilde{L}$ as arguments 
of $y$ and $\tilde{y}$ respectively,  is unchanged. 
The earlier treatment can be generalised by defining
\eq
y_k \equiv y_k(x,E,g,L)= \omega^k y(\omega^{-k}x,\omega^{-3M k}E,g
,\omega^{-3(M+1) k}L)~,
\label{defnewyk}
\en
so that
\eq
y'''_k  
- G (\frac{1}{x^2}\frac{d}{dx}-\frac{1}{x^3})y_k 
+ e^{-2 k \pi i }P(x,E,L) y_k =0 ~.
\label{difor}
\en
If we also define~$z_{k_1 k_2}$ as in~(\ref{zdef1}), then, for
$k_1$ and $k_2$ differing by an integer,
\eq
z'''_{k_1 k_2} -
  G (\frac{1}{x^2}\frac{d}{dx}-\frac{1}{x^3}) 
 z_{k_1 k_2}
- e^{-2 k \pi i }P(x,E,L) z_{k_1 k_2}  =0~, 
\label{auxdif3}
\en
(with $e^{-2 k_1 \pi i}=e^{-2 k_2 \pi i}\equiv e^{-2 k \pi i}$ )
which is the adjoint to~(\ref{difor}). Again we can recover
the original problem by shifting $k$ by a half-integer. As before, we
find that
$z_{-1/2,1/2}(x,E,g,L)= i \sqrt{3}\,y(x,E,g,L)$
and  (cf.~(\ref{tqqf}))
\eq
T(E,g,L)y_{-{1/4}} y_{{1/4}} = y_{-{1/4}}y_{{5/4}}
+ y_{-{3/4}} y_{{3/4}}+ y_{-{5/4}} y_{{1/4}}~.
\label{tqqfa2} 
\en
The presence of a non-vanishing $L$ has
however
introduced an extra complication.
To see this explicitly,  shift $k$ by $\pm 1/4$  to get  
\eq
T^{\pm} 
y_0  y_{\pm 1/2}= 
y_0  y_{\pm 3/2} 
+y_1  y_{-1/2}+ 
y_{-1}  y_{1/2}~~, 
\label{tpm1}
\en
with $T^{\pm}=T(E \omega^{\mp \fract{3 M}{4}},g, \mp i L) $. 
If this equation is rewritten in terms of the function $y(x,E,g,L)$,
both signs of $L$ appear:
for $k$ integer or half-integer we have 
\eq
y_{k} =  \omega^k y(x \omega^{-k} , E \omega^{-3M k},g,
 (-1)^{2k}  L)~.
\en 
Notice that the same does not happen for the argument $g$ (or $G$), 
which is why this problem did not arise before.
{}From an analytic point of view, $y(x,E,g,L)$ and  $y(x,E,g,-L)$
are just  two distinct  points of the same function, but
in the derivation of the nonlinear integral equation it is only
the analyticity in $E$ that is used.
To proceed, it is best to consider $L$ to be held fixed once and for all,
and to treat the pair of functions
$v_k = \omega^k y(\omega^{-k}x,\omega^{-3M k}E,g,L)$ and 
$\bar{v}_k = \omega^k y(\omega^{-k}x,\omega^{-3M k}E,g,-L)$ 
independently.
Then (\ref{tpm1}) becomes
\eq
T^{\pm} v_0 \bar{v}_{\pm 1/2}= 
v_0    \bar{v}_{\pm 3/2} +
v_1    \bar{v}_{-1/2}+ 
v_{-1} \bar{v}_{1/2}~~. 
\label{tpm3}
\en
This equation is very reminiscent of those given
in~\cite{PS} for the $A_2$-lattice model. There remains an
$x$-dependence
in (\ref{tpm3}) which can be eliminated, once again, by expanding
\eq
v =  D^{+}(E,g,L)~\chi_+ + 
D^{0}(E,g,L)~\chi_{0} +   D^{-}(E,g,L)~\chi_{-}~~, 
\label{specdet3}
\en
\eq
\bar{v}  =  \bar{D}^{+}(E,g,L)~\bar{\chi}_+  + 
\bar{D}^{0}(E,g,L)~\bar{\chi}_0 +   \bar{D}^{-}(E,g,L)~
\bar{\chi}_-~~, 
\label{specdet4}
\en
where  $\{ \chi_+, \chi_0, \chi_{-} \}$ and
 $\{ \bar{\chi}_+, \bar{\chi}_0, \bar{\chi}_{-} \}$
are alternative bases  defined via the behaviour near the origin
\eq
\chi_i(x,E,g,L)  \sim   x^{\lambda_i}+O(x^{\lambda_i+3})~, \qquad i=+,0,-\,,
\en
\eq
\bar{\chi}_i(x,E,g,L)  \sim   x^{\bar{\lambda}_i}+O(x^{\bar{\lambda}_i+3})~, 
\qquad i=+,0,-\,,
\en
and the  $\lambda_i$'s and  $\bar{\lambda}_i$' are respectively 
solutions of 
the indicial equations 
\eq
(\lambda-1)(\lambda (\lambda-2)-g(g+2))+L=0~~,~~
(\bar\lambda-1)(\bar\lambda (\bar\lambda-2)-g(g+2))-L=0~.
\label{lbal}
\en
If the  labeling is chosen consistently 
with  that of section \S\ref{dualitysec}, 
so that the $\lambda$'s and the $\bar{\lambda}$'s 
reduce to the quantities in~(\ref{lg}) when $L=0$, then 
\eq
T^{\pm} Q_0^{\pm} \bar{Q}_{\pm 1/2}^{\pm}= 
Q_0^{\pm}  \bar{Q}_{\pm 3/2}^{\pm}+ 
Q_1^{\pm}  \bar{Q}_{-1/2}^{\pm} +
Q_{-1}^{\pm}  \bar{Q}_{1/2}^{\pm}~~, 
\label{tpm4}
\en
with
\eq
Q^{\pm}(E,g,L)=E^{\fract{ \lambda_{\pm}-1}{3M}} 
D^{\pm}(E,g,L)~,~
Q^{\pm}_k = Q^{\pm}(\omega^{-3 M k}E,g,L)~,
\label{QDdef3}
\en
\eq
\bar{Q}^{\pm}(E,g,L)=E^{\fract{\bar{\lambda}_{\pm}-1}{3M}} 
\bar{D}^{\pm}(E,g,L)~,~
\bar{Q}^{\pm}_k  =
\bar{Q}^{\pm}(\omega^{-3 M k}E,g,L)~.
\label{QDdef4}
\en
This leads to two coupled sets of BA equations 
\bea
\prod_{k=1}^{\infty}  { E^{\pm}_k - \omega^{-3M}    E^{\pm}_n \over  
E^{\pm}_k - \omega^{3M}  E^{\pm}_n} &=& - 
\omega^{  2 \lambda_{\pm}- \bar{\lambda}_{\pm}-1}
\prod_{k=1}^{\infty}  { \bar{E}^{\pm}_k - \omega^{-3M/2} 
E^{\pm}_n \over  
\bar{E}^{\pm}_k - \omega^{3M/2} E^{\pm}_n}~\, , \\
\prod_{k=1}^{\infty}  { \bar{E}^{\pm}_k - \omega^{-3M}    
\bar{E}^{\pm}_n \over  
\bar{E}^{\pm}_k - \omega^{3M}  \bar{E}^{\pm}_n} &=& - 
\omega^{2 \bar{\lambda}_{\pm} - \lambda_{\pm}-1}
\prod_{k=1}^{\infty}  { E^{\pm}_k - \omega^{-3M/2} \bar{E}^{\pm}_n 
\over  E^{\pm}_k - \omega^{3M/2} \bar{E}^{\pm}_n}~\, .
\label{BA4}
\eea
Generalising the analysis of \S\ref{snie}, it should  be possible to derive
a nonlinear integral equation relevant to
this more general case. We expect 
that this
equation will coincide with the 
$a_2^{(1)}$-related case of the equations found in
\cite{Ma,ZJ}, in its
massless limit, but we will leave a detailed investigation
for future work.
%
%%%%
\resection{Conclusions}
We have continued to study the relationship
between
integrable quantum field theories
and ordinary differential equations, and in the 
process have obtained a novel nonlinear integral equation which is able to
describe the $\phi_{12}$, $\phi_{21}$ and $\phi_{15}$ perturbations of minimal
models within a unified framework. We have also found a natural generalisation
of the duality
symmetry enjoyed by the Schr\"odinger/massless sine Gordon 
system~\cite{BLZschr}.
A major theme has been that 
the $A_2$ structures hidden inside certain third-order ordinary 
differential equations,
and also inside certain integrable quantum field theories and BA 
systems, are very closely related. 
It seems clear that the correct way to generalise 
to yet further models is to look to differential equations of even
higher order. While this might appear to be a task of ever-increasing
complexity, there are some reasons to suppose that a more unified picture
will ultimately emerge.
ADE structures have been observed in many different, but 
related, settings in the context of integrable models
(see, for example,
\cite{BR,Zam2,Ma,ZJ,PED,RTV,Suz2} and references therein).
One might hope that the process of generalisation
will reveal similar phenomena
on the differential equations side of the correspondence, but
more case-by-case analysis will certainly
be required before this can be confirmed.

%%%%%%%%%%%%%%%%%%%%%%%%%%%%%%%%

\bigskip
\noindent
{\bf Acknowledgements -- }
We are grateful to
John Coleman, Clare Dunning,
Paul Fendley, Davide Fioravanti, Bernard Nienhuis and 
Francesco Ravanini  for useful discussions.
In addition,
RT thanks Durham University, and PED thanks the YITP, Kyoto, for 
hospitality during the final stages of the writing of this paper.
The work was supported in part by a TMR grant of the
European Commission, reference ERBFMRXCT960012; the visit of PED to
YITP was funded by a Daiwa-Adrian Prize.
PED thanks the EPSRC for an Advanced Fellowship, and
RT thanks
the Universiteit van Amsterdam for a post-doctoral
fellowship.

%%CHANGE:
\bigskip
\noindent
{\bf Notes added -- }

\noindent
(i) The `massless' nonlinear integral equation derived in \S3 has
appeared previously, in connection with the Izergin-Korepin model,
 in \cite{WBN}. 

\noindent
(ii) A conjecture due to Kausch et al~\cite{KTW,gabor} states that the
$\phi_{12}$ perturbation of ${\cal M}_{p,q}$ and the $\phi_{15}$ perturbation
of ${\cal M}_{p',q'}$ have identical ground-state scaling functions if (and
only if) $p'=p/2$, $q'=2q$. 
(This implies $p=2\,{\rm mod}\,4$, since $(p,q)$ and $(p',q')$ must both 
be coprime; such pairs are called `type II' in \cite{KTW}.)
It is easily checked that this equality follows from the recipe
for finding ground-state scaling functions given 
in \S6 above: the values of $M$ and $\alpha$ that
should be used in the two cases are identical, and so both are described
by the same nonlinear integral equation. We take this as additional
support both for our conjectures and for that of \cite{KTW}.

\noindent
(iii) In a recent paper~\cite{Suz3}, Suzuki has independently
remarked the relevance of
higher-order ordinary differential equations to integrable models
associated with the algebra $A_n$, though with a slightly different
emphasis from that adopted above.

\noindent
We would like to thank Ole Warnaar and the referee for bringing
ref.~\cite{WBN} to our attention, and Gabor Takacs for telling us
about the type II conjecture.

%%END OF CHANGE
%%%%%%%%%%%%%%%%%%%%%%%%%%%%%%%%
%

\end{document}